\documentclass[12pt]{JHEP3} 
\usepackage{amsfonts}
\usepackage{amssymb}
\usepackage{cite}
\def\be{\begin{equation}} \def\ee{\end{equation}}
\def\bea{\begin{eqnarray}} \def\eea{\end{eqnarray}} \def\ba{\begin{array}}
\def\ea{\end{array}} \def\ben{\begin{enumerate}} \def\een{\end{enumerate}}
  
\def\lll{\label}

 \title{  Romans type IIA theory and the heterotic strings}
\author{ Harvendra Singh
\thanks{email: singh@physik.uni-halle.de}
\\
 Fachbereich Physik, Martin-Luther-Universit\"at Halle-Wittenberg,\\
Friedemann-Bach-Platz 6, D-06099 Halle, Germany}

\abstract{
In this paper we study $T^2$  compactification
of six-dimensional massive type IIA supergravity  in presence of  
 Ramond-Ramond background fluxes. 
The resulting theory in four dimensions 
is shown to possess $SL(2,R)\times SL(2,R)\times O(4,20)$ duality 
symmetry.
It is shown  that specific elements of this  symmetry  relate 
massive type IIA compactified on $K3\times T^2$ (fluxes along $K3$) 
to the ordinary type IIA 
compactified on $K3\times T^2$ (fluxes along $T^2$) .
In turn, this relationship is exploited to relate Romans theory to 
heterotic strings.  
The D8-brane (domain-wall) wrapped on $K3\times T^2$ is found to 
correspond to  {\it pure gravity} heterotic
solution which is a direct
product of six-dimensional  flat space and a four-dimensional 
Taub-NUT instanton. 
}

\preprint{hep-th/0201206}
\keywords{ string, supergravity, compactification, duality}

\begin{document}
\newcommand{\eqn}[1]{(\ref{#1})}
\def\cC{{\cal C}} 
\def\cG{{\cal G}} 
\def\cd{{\cal D}} 
\def\a{\alpha}
\def\b{\beta}
\def\g{\gamma}\def\G{\Gamma}
\def\d{\delta}\def\D{\Delta}
\def\ep{\epsilon}
\def\e{\eta}
\def\z{\zeta}
\def\t{\theta}\def\T{\Theta}
\def\l{\lambda}\def\L{\Lambda}
\def\m{\mu}
\def\f{\phi}\def\F{\Phi}
\def\n{\nu}
\def\p{\psi}\def\P{\Psi}
\def\r{\rho}
\def\s{\sigma}\def\S{\Sigma}
\def\ta{\tau}
\def\x{\chi}
\def\o{\omega}\def\O{\Omega}
\def\k{\kappa}
\def\pa {\partial}
\def\ov{\over}
\def\br{\nonumber\\}
\def\ud{\underline}

\section{Introduction}

Recently a lot of attention is given to the
 studies of gauged and massive supergravity theories for the 
 reason that 
these theories play a significant role in AdS/CFT analysis 
\cite{maldacena} and also because they
might be useful in string phenomenology \cite{rsun}.
Massive supergravities are regarded as closed counterparts
of gauged supergravities. 
In massive theories some of the vector 
or tensor fields become massive upon eating other fields in 
the spectrum, analogous to
 a  Higgs type mechanism. In this procedure  the total degrees of
 freedom  remain unaltered and so do the number of supercharges, 
however it turns out that we need to sacrifice some of the global 
symmetries, like 
dualities. But it is not always so, certain fraction of these symmetries, 
at least the perturbative ones, could be restored. This is the goal of our 
investigation in  this paper to look for these unbroken global symmetries. 

A well studied and 
unique example of a massive theory of gravity is the massive type IIA 
supergravity 
in ten dimensions constructed by Romans \cite{roma}. 
In string theory, massive supergravities
can typically  be constructed through a  Scherk-Schwarz
reduction \cite{ss}, in which
 some field strength $dA_{(p)}$
 is given a non-trivial background value (flux) along the
compact directions \cite{berg1}. 
Established criterion for turning on such background fluxes 
with consistency is that the potential  $A_{(p)}$ 
appears only through its field strength $dA_{(p)}$ in the action, 
or equivalently in  the 
field equations. There has been several works 
along these lines in recent past
\cite{berg1,cow,blo,bs,mo,lavpop1,lavpop,bre,lupop, 
kkm,km,amees,hull}, for more latest works see \cite{ 
hls,janssen,CKLT,hs,micu,behrndt}. 

Amongst massive theories, massive type IIA
supergravities can be easily characterised solely by there field content 
in which the NS-NS 2-rank tensor field is massive.
In $D<10$ massive type IIA theories, in which tensor fields are 
massive, can also be obtained through generalised $K3$ and toroidal 
compactifications of the ordinary (massless) type IIA supergravities with 
Ramond-Ramond 
(RR) fluxes turned on \cite{hls,hs}. On the other hand massive heterotic 
supergravities with 
massive 1-form fields could be obtained by generalised Kaluza-Klein 
compactification \cite{bre,km}. Thus there is no way out we can make a  
connection 
between massive type IIA theories and massive heterotic strings because 
different type of fields carry masses \cite{hls},  though 
there is a duality relationship in $D=6$ involving ordinary 
type II and heterotic supergravities, for review see
\cite{sen}. 
Therfore  
unless there is a machenism by which we can trade massive 2-rank 
tensor fields
 into a massive vector fields we cannot help us out.
Fortunately there is such a relationship in four dimensions where a 
massive tensor field carries same degrees of freedom as a massive vector 
field 
carries \cite{quevedo}. So in four dimensions it would be  possible to 
dualize a 
massive 2-rank tensor field into massive 1-form field.
Hence the study of massive theories in $D=4$ is very crucial for 
duality relationship between massive type 
IIA and heterotic strings. Though, we will not show this duality 
relationship
between massive tensors and massive vector fields 
explicitely, we shall be adopting an equivalent tool 
of compactifications and use duality symmetries, which do this job 
implicitely.
 
Paper is organized in the following way.  We work out a
generalized $T^2$  compactification
of the six-dimensional  massive type IIA supergravity  \cite{hls}
with RR 2-form fluxes turned on.
Our goal is to investigate the fate of the
perturbative $SL(2,R)\times SL(2,R)\times O(4,20)$ 
duality symmetry.
We  then are interested in  relating this $6D~ N=2$ massive type
IIA compactified on $T^2$  to the ordinary $6D~ N=2$ type IIA compactified 
on $T^2$ with fluxes and eventually 
relate it to heterotic string theory compactified on $T^4$.
In section~2 we briefly recall $6D$  massive type IIA sugra
and work out its compactification  on $T^2$ with RR fluxes.
We discover that provided RR 2-form fluxes are 
turned on the resulting four-dimensional
theory can be presented in a manifestly $SL(2,R)\times SL(2,R)\times 
O(4,20)$ covariant 
form. In section~3 we provide a mechanism to relate Romans 
theory 
compactified on $K3\times T^2$ 
(with fluxes)  to the ordinary type IIA compactified on $K3\times T^2$ 
(with fluxes) which is in turn related to heterotic strings.   
In section~4 we study the vacuum solutions
of this massive type IIA supergravity and 
relate them to the solutions of ordinary IIA
by using the elements of duality group. We then use standard 
heterotic-type IIA duality in six dimensions to relate corresponding 
ordinary IIA vacua to heterotic solutions. Particularly the 
D8-brane solution is shown to correspond to
a {\it pure gravity} heterotic vacua in ten dimensions which contains a
domain-wall-type instanton line element. 
Thus perturbative duality symmetry
relates the vacua of massive type IIA and ordinary type IIA theories 
in six dimensions provided we have $T^2$ isometries. 
This property also allows us to further relate massive type IIA vacua
to six-dimensional heterotic solutions.
We have summarized our results in the section~5.

\section{Toroidal compactification}
The compactification of Romans type IIA supergravity \cite{roma} on $K3$ 
with RR fluxes was provided in \cite{hls}. The resulting bosonic  
composition of $6D$ $N=2$ 
massive type II theory is given by 
\begin{eqnarray} \label{massive2a}
S_6&=&\int \bigg[ e^{-2\hat\f}\left\{\hat
R~{\hat\ast}1+4~\hat d\hat\f{\hat\ast} \hat d\hat\f -
{1\over2}\hat H_{(3)} {\hat\ast}\hat H_{(3)} 
+{1\over8}Tr \hat d\hat M\hat\ast \hat d\hat M^{-1}\right\}
-{1\over2}\hat F_{(2)}^a \hat\ast \hat M^{-1}_{ab}\hat F_{(2)}^b \br && 
-{1\over2} m^a {\hat\ast}\hat M^{-1}_{ab}m^b 
 + {1\over2} \hat B \hat F^a \hat F_a -{1\over 2} \hat B^2m^a\hat F_a
+{1\over 3!} \hat B^3m^am_a 
\bigg]\ ,
\lll{1a}
\end{eqnarray}
 where we have adopted the notation that 
 every product of forms is understood 
 as a wedge product.\footnote{
 The signature of the metric is $(-+\ldots+ )$
 and  for a $p$-form  we use the convention
 $F_p={1\over
 p!}F_{M_1...M_p}dx^{M_1}\wedge...\wedge dx^{M_p}\ ,$ 
 while the Poincare dual is given by
 ${\hat\ast}F_{p}={1\over p!(6-p)!} F_{M_1...M_p}
\epsilon^{M_1...M_p}_{~~~~~~~M_{p+1}...M_{6}}~dx^{M_{p+1}}\wedge...\wedge
 dx^{M_{6}} ,~
 ~{\hat\ast \hat\ast}F_{p}=-(-1)^{p(6-p)}F_p\ ,$
 and $~{\hat\ast} 1 =\sqrt{- \hat g}\, [d^{6}x]$.} 
Here $\hat\f$ is the dilaton field, $\hat B_{(2)}$ is NS-NS tensor 
field and 
$\hat A^a~ (a=1,\cdots,24)$ are RR vector fields which form a vectorial 
representation of the duality symmetry group $O(4,20)$. $m^a$ represents 
24 mass (or flux) 
parameters  which also form a vector of $O(4,20)$. The  field strengths in 
the action 
\eqn{1a}
are given by
\be
\hat H_{(3)}=\hat d\hat B_{(2)}\ ,
\qquad 
\hat F_{(2)}^a= \hat d\hat A_{(1)}^a +m^a\hat B_{(2)}\ ,
\qquad
\lll{a22}\ee
There are eighty scalar fields those parameterize 
$O(4,20)/O(4)\times 
O(20)$ coset matrix $ \hat M^{ab}$ which 
satisfies
\be
 \hat M^T L~ \hat M =L,~~~  \hat M^{-1}=L~ \hat M~ L,~~~\hat M^T=\hat M
\ee
where $L$ is $O(4,20)$ metric. Thus  above six dimensional action has 
manifest $O(4,20)$ invariance \cite{hls}. Once  $m^a$ is vanishing, above 
action represents ordinary type IIA compactified on $K3$.

We shall now work out the compactification of above theory on $T^2$ in 
presence of RR fluxes. To recall,
 $T^2$ compactification of Romans  type IIA theory has been 
worked out in detail in \cite{hs}, so we shall omit various fine details 
here. 
Thus for the 
6-dimensional sechsbein we  take the standard toroidal ansatz
\be\label{metric}
\hat e^{\hat a}_{M}\, (x,y)= \left( \ba{cc}e^a_\m(x)&~~~e^i_m K^m_\m\\
0&e^i_{m}\ea\right)\ ,
\ee 
where coordinates $y^m,~(m=1,2),$ are tangent to the tori. 
The internal metric on tori
is  given by $G_{mn}(x)= e_m^i \d_{ij}e_n^j$ while
the 4-dimensional spacetime metric is $ g_{\m\n}(x)= e_\m^a 
\eta_{ab}e_\n^b$. 
$K^m_\m$ are the
Kaluza-Klein gauge fields and we define
1-forms  $\eta^m=dy^m+K^m_{(1)}$.

The standard toroidal ans\"atze for
 the dilaton, tensor field and the moduli matrix are taken to be
\be
\hat\f(x,y)=\hat \f(x)\ ,~~
\hat B_{(2)}(x,y)=\bar B_{(2)}(x)+ \bar A_m(x)~ \e^m+{1\over 
2}b_{mn}(x)~\e^m 
 \e^n\ ,~~  \hat M(x,y)=M(x)\ ,
\ee
where $\bar B_{(2)}= B_{(2)}-{1\ov2}\bar A_m K^m$ with $ B$  being 
a 2-form in four spacetime
dimensions, $\bar A_m$ are two vector
fields      
and the $b_{mn}$ represents scalar fields antisymmetric in $m,n$ 
indices. With these ans\"atze the NS-NS part of the action
 \eqn{massive2a} reduces
to \cite{ms}
\bea
&&\int  
  e^{-2\f}\bigg[
R~{\ast}1+4~d\f{\ast} d\f -
{1\over2} H {\ast} H 
+{1\over4}dG^{mn}\ast dG_{mn}-{1\ov4}db_{mn}\ast db_{pq}G^{mp}G^{nq}  
\br &&-{1\over 2}dK^m\ast dK^nG_{mn}  
-{1\ov 2}(d\bar A_m-b_{mp}dK^p)\ast (d\bar A_n-b_{nq}dK^q)G^{mn}
\br &&+{1\over8}Tr d M\ast d M^{-1}
\bigg]~,
\eea
with
\be
\label{deff1}
2\f=2\hat\f-{1\ov2} \ln~det (G_{mn})  ,\qquad H=d B -{1\ov2}
(\bar A_m dK^m+K^md\bar A_m)\ . 
\ee
Next, for twenty-four 1-forms $\hat A^a$ 
we would consider a generalized 
Kaluza-Klein ansatz where RR fluxes along $T^2$ are included.\footnote{From 
here onwards we switch to a notation where an
 $O(4,20)$ vector like $A^a$ is represented simply by an 
overhead vector, like $\vec A$.} 
This generalization is possible since  
$\hat{\vec A}$  appear in the action \eqn{1a} 
only through derivatives, therefore an appropriate background value
can be consistently turned on. 
We take  an ansatz
$ {\hat{\vec A}}(x,y)=  \vec A_{(1)}(x)+  \vec A_{(0)m}(x,y) dy^m ,$ 
where scalar fields $\vec A_{(0)m}(x,y)$ are allowed to retain a  
dependence on the coordinates of the torus. The consistency of toroidal 
reduction requires it can  at most be
a linear dependence on the torus coordinates, 
we define
 \be
\vec A_{(0)m}(x,y)=\vec a_m(x)-{1\ov2}\vec w_{mn} y^n,
\lll{ansatzz}\ee
where new constants $\vec w_{mn}$ are
antisymmetric in indices. This  gives us
\bea
&&{ \hat d} {\hat{\vec A}}= \cd \vec A + \cd \vec a_m \eta^m +{1\ov 2}\vec 
w_{mn} \e^m\e^n,
\lll{ansatz1a}
\eea
where various $\cd$-derivatives are defined as 
\bea
&& \cd \vec A= d\vec A -d\vec a_m K^m +{1\ov2}\vec w_{mn}K^mK^n~,~~~ \cd 
\vec a_m=d \vec a_m + \vec w_{mn} 
K^n\ .
\eea
Note that various forms are distinguishable by the symbols and 
the internal indices they carry. Thus
through above  generalized ansatz we have effectively 
introduced  24 new  parameter $\vec w_{12}$ in 
the form of fluxes. 
This generalization has been possible only because  
$\hat A$'s appears in the action 
 covered with derivative and $T^2$ has  2-cycle along which
 an appropriate background flux
could be turned on. 

After compactification the bosonic spectrum of  four-dimensional
theory consists of the graviton $g_{\m\n}$, dilaton $\phi$, 2-form
$ B$, 4 scalars from the components of metric and tensor field, and 4 
 1-forms $(\bar A_m,~K^m)$ in the NS-NS
sector. From the R-R sector we have $2\times 24$ scalars $\vec a_m$, 
 and twenty-four 1-forms $\vec A$ whose field strength are 
(anti)self-dual in $D=4$. 
Also we have two sets of 24 constant parameters in the form of fluxes 
$\vec w_{12}$ and 
masses $\vec m$.
This is the precisely the bosonic field
content of $N=4$ type II supergravity theory in $D=4$.
In the massless case these fields fit in various representation of the
T-duality group $SL(2,R)\times SL(2,R)\times O(4,20)$.
Here too various fields combine into the   
$SL(2,R)\times SL(2,R)\sim SO(2,2)$ representations  as
\bea
&&
A^{ru}_{(1)}=(A^{r1},A^{r2}),~~A^{ru=1}=(\bar 
A_1,~\bar A_2),~~A^{ru=2}=({K^2},~~-{K^1}),\br
&&
\vec a^r_{(0)}=(\vec a_1, \vec a_2),~~,~~\vec m^u=(-\vec w_{12},~\vec m),
\eea
where indices $r=1,2$ belong to first $SL(2,R)$ 
while indices $u=1,2$ belong to the second $SL(2,R)$ group.
Note that the mass and flux parameters also fit into a 
fundamental representation of $SL(2,R)$.

In order to obtain the action  involvng low energy 
modes of this theory we
substitute the ansatz \eqn{metric}-\eqn{ansatz1a} into the action
\eqn{1a}. 
The resulting four-dimensional bosonic action reads in the kinetic part
 
\bea
S_{4}&=&\int\bigg[e^{-2\f}\bigg\{ R~\ast 1+4~ d\f \ast d\f -
{1\ov2} H \ast H-{1\ov2} dA^{ru}\ast dA^{sv}
N^{-1}_{rs}{\cal M}^{-1}_{uv} + \br 
&&
+{1\over 4} {\rm Tr}~ d{\cal M}^{-1} \ast
d{\cal M}+ {1\over 4} {\rm Tr}~ d{N}^{-1} \ast
d{N} +{1\over8}Tr d M\ast d M^{-1}\bigg\}
-{1\over2} {\vec F}_{(2)}\ast { M}^{-1} 
{\vec F}_{(2)}~\sqrt{G} \br &&
-{1\over2} {\vec S}_{(1)}^r M^{-1}\ast 
{\vec S}_{(1)}^s   N^{-1}_{rs}
-{1\over2} \vec m^u M^{-1}\ast \vec m^v{\cal M}^{-1}_{uv}~ 
\bigg]+S_{CS}~,
\lll{action8}
\eea
where the Chern-Simon part of the action is
\bea
&&S_{CS}= \int {\ep^{mn}\ov2}\bigg[
 b_{mn} \vec F L \vec F +\vec F L \left\{ \bar B\vec w_{mn} -2 \bar 
A_m\cd\vec a_n+\vec m \bar A_m \bar A_n \right\}- {1\ov2}\bar B^2\vec m 
L \vec w_{mn}\br
&&~~~~~~~  -\bar B (d\vec a_m+\vec w_{mp}K^p)L(d\vec a_n+\vec w_{nq}K^q)  
\bigg]~.
\lll{action88}
\eea
Various  field strengths in the above action are 
\be
 H_{(3)}=d B + 
A^{ru}dA^{sv}\eta_{rs}{\cal L}_{uv}\ ,~~
{\vec S}_{(1)}^r=d\vec a~^r+ \vec m_u~A^{ru}~,~~
{\vec F}_{(2)}=\cd \vec A+ \vec m (B- {1\ov2} \bar A_mK^m)~.
\lll{action88a}
\ee
The indices $r$ and $u$ can be raised or lowered by the use of 
two $SL(2,R)$ metrics $\eta$ and ${\cal 
L}$, respectively, which  are given by 
$$ \eta_{rs}=\left(\ba{cc} 0 &-1\\ 1 &0\ea\right)\equiv {\cal L}_{uv}\ . 
$$
The uni-modular matrices which belong to two  
$SL(2,R)/SO(2)$  
cosets are given by
\be
N^{-1}=\sqrt{G}\left(\ba{cc} G^{11} &~~G^{12}\\ G^{21} 
&~~G^{22}\ea\right),~~~~{ \cal M}^{-1}= 
{1\over \sqrt{G}}\left(\ba{cc}  
1 &~~~-b\\ -b &~~~(b)^2+G\ea\right),
\ee
and they satisfy  $N^T \eta N=\eta,~~{\cal M}^T {\cal L}{\cal 
M}={\cal L}~.$
We shall take $\ep^{12}=1$ and we have defined $b_{12}\equiv b$, 
$\vec w_{12}\equiv \vec w$.  
Under the $SL(2,R)$ transformations which act upon $u,v$ indices
\bea
&& g_{\m\n}\to g_{\m\n},~~~\f\to\f,~~~B\to B~,~~~M\to M~,\br
&&{\cal M}\to \L{\cal M}\L^T~, ~~~~
{ A}^{ru}\to \L^u_{~v}{ A}^{rv}~,~~~~ \vec m^u\to\L^u_{~v}\vec m^v~,
\lll{trans}
\eea 
 $ \L^T~{\cal L}~\L={\cal L}$, alongwith the transformation for 
$\vec A$ which is yet to be determined. The
 second $SL(2,R)$ group acts only on $r,s$ 
indices in a similar way. 
The action \eqn{action8} has manifest $O(4,20)$ invariance from 
the beginning 
as all $O(4,20)$ indices are contracted with metric $L$.

Note that the kinetic terms in the action \eqn{action8} except the terms 
involving 24 two-form field strength ${\vec F}$ 
remain invariant under the action of above T-duality group. 
It remains to be seen if the field equations 
and the Bianchi identity for 1-form 
potential $\vec A~$   transform covariantly.
From the 4-dimensional action in \eqn{action8}  the field equations for 
$ \vec A$ are  
\bea
&&d\left(-\sqrt{G}\ast M^{-1}\vec F + b~L \vec F + {1\ov2}\ep^{mn}L\big[
\vec w_{mn}\bar B -2\bar A_m 
\cd\vec a_n+ \vec m \bar A_m\bar A_n 
\big]\right)=0~,
\lll{a2a}
\eea
while the Bianchi Identities are
\be
d\left( \vec F+ d\vec a_mK^m-{1\ov2}\vec w_{mn}K^mK^n-\vec m\bar 
B\right)=0~.
\lll{a2a1}\ee
We now define  the dual field strengths as $ \tilde{\vec F}= -\sqrt{G}\ast 
ML\vec F + b~ \vec F$, 
 then 
field equations \eqn{a2a} and  the  Bianchi 
identities for ${ \vec A}$ form a $SL(2,R)$ covariant set of equations
provided $\vec F$ and its dual transform as a vector under $SL(2,R)$ 
transformations. Thus $SL(2,R)$ symmetry mixes $\vec F$ with its dual. 
One  can also combine $\vec F$ and its dual into a $SL(2,R)$ field 
strength 
\bea
\vec F^u&=& d\vec{\cal A} ^u - d\vec a_r A^{ru} - {1\ov 2}\vec 
m_vA^v_sA^{su}+ \vec m^u  B~,
\eea
where  $ \vec{\cal A}^u=({\tilde{\vec A}}, {\vec A})$.

This completes our analysis of the four-dimensional massive type II 
supergravity action which we have 
shown to possess an explicit $SL(2,R)\times SL(2,R)\times O(4,20)$ 
duality symmetry at the level of field equations provided 
the fluxes on $T^2$, $\vec w$, and the masses, $\vec m$, which come 
from $K3$-compactification,  transform as 
$SL(2,R)$ doublet. 

The action  \eqn{action8}  possesses Stueckelberg gauge invariances 
\cite{roma} 
which is obvious from the investigation of the field strengths in 
eq.\eqn{action88a}. Through these gauge invariaces, the vector fields 
$A^{ru}$ can eat the scalars $a^r$ and can become massive. Similarly the 
tensor field $ B$ can eat one of the two vector fields ${\cal A}^u$ and 
can become massive. However, this process of swellowing in of the fields 
will break the duality symmetry explicitely. 

\section{Massive IIA and heterotic strings}
 
As we have seen in the previous section, the restoration of the  T-duality 
symmetry of the massive II theories in $D=4$ has been  a direct 
consequence of our generalised flux-type  
ansatz in \eqn{ansatzz}. This tells us that in this framework a wide class 
of type II theories with various RR
fluxes on $K3$ and/or $T^2$, in fact, get 
unified. Specifically under the $SL(2,R)$ elements of 
this duality symmetry the 
$K3$-masses (or fluxes) and the 
$T^2$-fluxes are mixed up and rotated. We now discuss a particular case 
which is of  interest in the rest of this paper. 
Consider a compactification of $6D ~N=2$ massive II theory \eqn{1a} 
on $T^2$ without fluxes (i.e. $\vec w=0$), 
the compactified four dimensional massive theory will then be characterisd 
by the mass vector 
$m^u=(\vec 0,~\vec m)$.\footnote{Here $\vec 0$ reprsents 24-dimensional 
vector with zero entries.} Similarly if we compactify a $6D ~N=2$ 
ordinary type IIA theory (that means $\vec m=0$ in \eqn{1a}) on 2-torus with 
fluxes, the resulting  four-dimensional theory will be chracterised by 
a different mass vector  $m^u=(-\vec w,~\vec 0)$. These two 
four-dimensional 
theories obtained in two different ways can
simply related by the following $SL(2,R)$ element
\be\label{U3}
\L=\left(\ba{cc} 0&-1\\  1&0 \ea\right)\ .
\ee 
In many ways  this is analogous to the 
identification between 2-form RR flux and the Romans' mass parameter
when  $10D$ massive IIA is 
compactified  on $T^2$ and a ordinary type IIA compactified  on 
$T^2$ with RR flux \cite{hs}. 
Other elements of above $SL(2,R)$  group mix two types of fluxes which 
could be used to generate new background configurations. In the rest of 
this paper our 
aim is to relate massive type IIA backgrounds 
to  heterotic string backgrounds.  

Having achieved this relationship between $6D$ $N=2$ massive IIA 
theory and the ordinary $6D$ $N=2$ ordinary type IIA, both compactified on 
2-torus, former without fluxes and the latter with fluxes, it is now 
straight forward 
to achieve the heterotic connection via following six-dimensional
S-duality relations between type IIA compactified on $K3$ and heterotic 
string theory compactified on $T^4$
\cite{sen}
\bea
&&\f_{het}= -\f_{II},~~~g_{\m\n}^{het}=e^{-\f_{II}}g_{\m\n}^{II},~~~
M_{het}=M_{II},~~~\vec A_{het}=\vec A_{II},
\br && H_{(3)}^{het}=e^{-\f_{II}} \ast H_{(3)}^{II}~. 
\lll{sdual}
\eea 
 
In this approach Romans theory compactified on  $K3(fluxes)\times T^2$ 
is mapped to the ordinary type IIA compactified on $K3\times 
T^2(fluxes)$ which in turn is related by duality \eqn{sdual} to heterotic 
string compacified on $T^4\times T^2(fluxes)$.
In the next section we shall take an explicit examples of domain-wall  
solutions and display 
this duality chain.

\section{D8-brane vs heterotic instanton}

The ten-dimensional massive IIA supergravity theory has  D8-brane
(domain-wall) solutions which preserve sixteen supercharges \cite{berg1}. 
In the string frame  metric this solution is given by
\bea
&&d s^2_{10}= H^{-1/2} (-dt^2 + dx_1^2+\cdots +dx_8^2  
) +H^{1/2}dz^2,
\br &&2\f_{10}=-{5\ov2}\ln H,
\lll{wall}
\eea
where 
$H=1 +{m}|z-z_0|$ is a harmonic function of only the transverse
coordinate $z$ and all other fields have vanishing background values, $z_0$
refers to the location of the domain-wall.
We  compactify this solution  on $K3$ by 
wrapping four of its world-volume directions, say $x_5,..., x_8$.  
The corresponding  six-dimensional domain-wall  solution 
of the action \eqn{1a} can be  written down as \cite{hls}
\bea\label{sixss}
&&d\hat s_6^2= H^{-1/2} (-dt^2 + dx_1^2+dx_2^2+dx_3^2+dx_4^2) 
+H^{1/2}dz^2\ ,
\br &&2\hat\f_6=-{3\over 2} \ln H \ ,
~~~{\hat M}={\rm 
diag}(H^{-1},1,1...,1,1, H)\ , 
\eea
with mass vector $\vec m=(0,...,0,m)$.\footnote{The last  entry in 
the mass vector 
$\vec m$ represents the mass parameter of Romans theory.} 
Clearly this vacuum configuration corresponds to the situation when 
there is no background flux along $K3$.
The solution (\ref{sixss}) is left with  $8$ unbroken supersymmetries. 
Further compactification of this on $T^2$ gives us a solution of the 
action \eqn{action8}
\bea\label{sixs}
&&d s_4^2= H^{-1/2} (-dt^2 + dx_1^2+dx_2^2) 
+H^{1/2}~dz^2 \ ,
\br &&2\f_4=- \ln H \ ,~~~{ M}={\rm 
diag}(H^{-1},1,1...,1,1, H)\ , \br
&&\vec m^u=(\vec 0, \vec m),~~~~{\cal 
M}={\rm diag}(H^{-{1\ov2}},H^{1\ov2}),~~~
{N}=I_2. 
\eea

Now, by applying $SL(2,R)$ 
transformations \eqn{trans} on the  fields in \eqn{sixs}
the  solutions with non-trivial R-R fluxes can be generated.
Let us consider the specific case where
 $SL(2,R)$ transformation is given by \eqn{U3}.
Inserting $\L$ and the configuration \eqn{sixs} in \eqn{trans} we
get
\be
\vec m^u=\left(\ba{c} \vec 0\\ \vec m\ea\right) \to \vec m'^u=
\left(\ba{c} -\vec m\\ \vec 0\ea\right),~~~
M\to M,~~N\to N,~~{\cal M}\to{\cal M}^{-1},
\lll{dual1}
\ee
while four-dimensional metric and the dilaton  remain invariant. The
transformed mass vector $\vec m'^u$ implies that the new
configuration is a solution of a $6D$ ordinary IIA
compactified on $T^2$   
with 2-form fluxes given by $\vec w=(0,....,0,m)$.
Lifting the  rotated solution \eqn{dual1}  to six dimensions, 
we get the following ordinary type IIA
configuration (we write new fields  with a prime)
\bea\label{news}
&&d \hat s_6^{'2}= H^{-1/2} (-dt^2 +
dx_1^2+dx_2^2)  + H^{1/2}(dz^2 + dy_1^2+dy_2^2 )\ ,
\br &&2\hat\f'_6=-{1\ov2} \ln  H\ ,\qquad  
{\vec {\hat A}}{'}=\vec w~ y_{[1}dy_{2]}~ ,\br
&&~~~{\hat M}'={\hat M}={\rm diag}(H^{-1},1,1...,1,1, H)\ .
\eea
This is in accordance with our ansatz in \eqn{ansatzz} and corresponds to 
swiching on the flux. 
 Since this solution is obtained by incorporating T-duality rotation
 \eqn{trans} the number of preserved supersymmetries will remain 
unchanged. 
Thus by making an $SL(2,R)$  transformation
we have transformed domain-wall solution \eqn{sixss} of 6D massive IIA 
theory 
into a domain-wall solution \eqn{news} of $6D$ ordinary type IIA 
which is supported by a non-trivial 2-form flux.
Thus, the  four-dimensional perturbative duality
interpolates between vacua of massive type IIA  
and ordinary type IIA. It  is parallel to 
the situation encountered 
in the case of massive type II duality in
$D=8$ \cite{hs}.\\ 

\noindent{\it Heterotic instanton:}
Since ordinary  type IIA theory on $K3$ 
is equivalent to heterotic theory on 
$ T^4$,  in order to relate  solutions of $6D$ massive IIA theory  to 
six dimensional heterotic string vacua we need first to map
them to the vacua of ordinary IIA by using the $SL(2,R)$ element 
\eqn{U3} and then use the relations \eqn{sdual}. Let us  consider for 
definitness the configuration in 
eq.\eqn{news} which is already a $6D$ ordinary IIA
background and  can therefore be mapped to heterotic side using 
\eqn{sdual}.
After, some straight forward calculation we get the following 
six-dimensional ordinary
heterotic solution (in string frame)
\bea\label{M5a}
&&ds_{6,het}^2= 
  -dt^2 +dx_1^2+dx_2^2  + H(dz^2 + dy_1^2+dy_2^2 )\ ,\br
&&2\f_{6,het}={1\over 2} \ln 
H~,~~~M_{het}=M_{II}=diag(H^{-1},1,...,1,H)~,\br
&&   
{\vec A}_{het}=\vec w~ y_{[1}dy_{2]}~,~~~\vec w=(0,...,0,m)~,
\eea
where the harmonic function $H=1+m|z-z_0|$. Note that heterotic 
string theory compactified on $T^4$ has the T-duality group $O(4,20)$ and 
 ${\vec A}_{het}$ belongs to the vector representation of this group 
\cite{ms}. 
It could be easily seen that the vector field  in \eqn{M5a} 
corresponds to a constant
field strength in $y_1,y_2$ directions which are along $T^2$.  A  
compactification of heterotic strings along these 
coordinates with such 
background fluxes gives rise to masses in four dimensions, see 
\cite{bre,km}.  On the other hand when \eqn{M5a} is oxidised to  ten 
dimensions, as in 
ordinary toroidal cases, we 
obtain following 10D heterotic vacua
\bea\label{M5}
ds_{10,het}^2&=&  H^{-1}[d\tau+{m\ov2}(y_1dy_2-y_2dy_1) ]^2 
+H(dz^2+dy_1^2+dy_2^2)- dt^2 +
\sum_{i=1}^5dx_i^2 \ ,\br
\eea
where $\tau$ is one of the coordinates along $T^4$ on 
which heterotic string is compactified.  
$y_1$ and $y_2$ are also periodic but are along $T^2$. This pure 
gravity heterotic 
vacua  preserves only 8 supersymmetries and 
 has the geometry which is a product 
of a 4-dimensional Taub-NUT instanton, ${\cal E}^4$, and a 
6-dimensional  Minkowski space, $M_6$. Properties of these Taub-NUT type 
instanton line element ${\cal E}^4$ are discussed in detail \cite{gibbons, hs}.

Thus D8-brane wrapped on $K3\times T^2$  emerges 
from purely geometrical 
configuration of the 
heterotic strings such that it involves  `domain-wall-instanton' \eqn{M5}. 
Compare \eqn{M5} with M-theory instanton which is also related via duality 
to the D8-brane wrapped on $T^2\times S^1$ and is given by 
\cite{hs}\footnote{
 11-dimensional solutions similar to  
\eqn{M55} originally appear in  \cite{cow,lavpop}. The line element in 
\eqn{M55} differs  only in 
the structure of 
the vector field  from  previous occasions.} 
\bea\label{M55}
ds_{11}^2&=&  H^{-1}[dx_{11}+{m\ov2}(y_1dy_2-y_2dy_1) ]^2
+H(dz^2+dy_1^2+dy_2^2)- dt^2 +
\sum_{i=1}^6dx_i^2 \ ,\br
&& H=1+m|z-z_0|\ ,    
\eea
where $x_{11}$ is the coordinate of 11-dimensional circle $S^1$. This 
solution however preserves 16 supersymmetries \cite{hs}. This is quite 
consistent as  Heterotic theory is obtained by orbifolding of M-theory.
These triad of solutions 
\eqn{wall}, \eqn{M5} and \eqn{M55} thus represent the same duality web 
involving $Type IIA\longleftrightarrow M-theory \longleftrightarrow 
Heterotic$  chain.
It has been shown   \cite{hs} that
 M-theory compactifications on  $M_7\times 
{\cal E}_4$  correspond to $10D$ massive IIA  compactification on 
2-torus.  Here we have presented an evidence that M-theory 
compactifications on $Z_2$ orbifold of $M_7\times 
{\cal E}_4$ should corresspond to heterotic compactification on $M_6\times 
{\cal E}_4$.

\section {Summary}
To summarize in this work we have studied the  
$T^2$ compactification of 
six-dimensional massive type IIA theory \cite{hls} 
with  Ramond-Ramond background fluxes corresponding to 2-form field 
strength. 
We have found that the resulting four-dimensional theory has 
$SL(2,R)\times SL(2,R)\times O(4,20)$ global symmetries, same as the 
perturbative duality symmetries which appear in ordinary 
compactifications.
The mass and  flux parameters transform under $SL(2,R)$ accordingly.
Thus the perturbative T-duality survives  at
the massive level, though in a different form that it requires appropriate 
masses and fluxes  to be switched on. 
Next we have shown that the elements of this duality symmetry
relate  
Romans theory compactified on $K3\times T^2$ (with RR fluxes along $K3$) 
with  ordinary type 
IIA  compactified  on $K3\times T^2$ (with RR fluxes along the  
$T^2$). 
This relationship between ordinary and massive IIA theories 
compactified  on $K3\times 
T^2$ fluxes has
led us to provide a heterotic string interpretation for massive IIA 
theory. 
As an example we have shown that the 
wrapped D8-brane solution of massive
type IIA turns out to be $SL(2,R)$ dual
of the solution of ordinary type IIA theory with flux,
which in turn is related to {\it pure gravity} vacua of heterotic 
string theory which is a
direct product of $6D$ Minkowski spacetime and a $4D$ 
 Ricci-flat instanton. The instanton line element is a domain-wall
generalization of  Taub-NUT instantons 
\cite{gh}. We 
recall that in \cite{hs} we have  shown that D8-brane are 
also related to the compactifications of M-theory invloving such
instantons and these solutions  have 16 supersymmetries intact. 
While the Heterotic 
solution \eqn{M5} has only 8 supercharges intact. This is 
entirely consistent given  
the fact that heterotic strings are orbifolds of M-theory.

\acknowledgments
I am grateful to  J. Louis for many interesting discussions and for 
carefully 
reading the draft of this paper. I  would also like to thank  A. Micu 
for useful discussions. 
This work is supported by AvH (the Alexander von Humboldt foundation).

\end{document}